# Using a Convolutional Neural Network and Explainable AI to Diagnose Dementia Based on MRI Scans

Tyler Morris, Ziming Liu, Longjian Liu and Xiaopeng Zhao

*Abstract*— As the number of dementia patients rises, the need for accurate diagnostic procedures rises as well. Current methods, like using an MRI scan, rely on human input, which can be inaccurate. However, the decision logic behind machine learning algorithms and their outputs cannot be explained, as most operate in black-box models. Therefore, to increase the accuracy of diagnosing dementia through MRIs, a convolution neural network has been developed and trained using an open-source database of 6400 MRI scans divided into 4 dementia classes. The model, which attained a 98% validation accuracy, was shown to be well fit and able to generalize to new data. Furthermore, to aid in the visualization of the model's output, an explainable AI algorithm was developed by visualizing the outputs of individual filters in each convolution layer, which highlighted regions of interest in the scan. These outputs do a great job of identifying the image features that contribute most to the model's classification, thus allowing users to visualize and understand the results. Altogether, this combination of the convolution neural network and explainable AI algorithm creates a system that can be used in the medical field to not only aid in the proper classification of dementia but also allow everyone involved to visualize and understand the results.

*Clinical Relevance*— This model promises to be a helpful diagnostic tool through a 98% accurate classification of dementia and the use of explainable AI to show regions of interest in the MRI scan.

## I. INTRODUCTION

Dementia, which includes conditions such as Alzheimer's Disease and Parkinson's, is a neurodegenerative disease that can cause physical disability, cognitive disorders, and most notably memory loss [1]. The latest report from the Alzheimer's Association indicates that, in 2023, 6.7 million Americans were affected by dementia, and that number is expected to rise to 12.7 million by 2050 [1]. As such, the need to diagnose the condition is on the rise, especially since an early diagnosis lessens a person's cognitive decline [1].

One reliable method of diagnosing dementia is through a magnetic resonance imaging (MRI) scan of the brain. These scans can show regions of the brain that are degraded due to dementia, putting a tangible assessment of a condition that is otherwise expressed primarily in cognitive symptoms [2]. Normally, a trained professional, such as a neurologist or radiologist, would examine the MRI and make a diagnosis. However, there are times were this process can take weeks due to specialists being unavailable to make consultations [3].

The human eye can also make mistakes since different stages of dementia can be very similar in an MRI scan [2]. So even though MRI scans are one of the most conclusive ways to diagnose dementia, it is necessary to make improvements to ensure its accuracy and speed.

To address the issues with the MRI diagnosis of dementia, several researchers have attempted to use artificial intelligence (AI) and machine learning to create algorithms that can read an MRI and classify it as either demented or non-demented [2-5]. However, this method has a few problems that need to be addressed. First, dementia is a progressive disease with several stages, and each stage comes with its form of treatment [1]. As such, any AI algorithm used to detect dementia in an MRI scan must be able to classify the stage of the disease so the person can get the proper treatment. Furthermore, many AI models, like neural networks, are so-called "black boxes", which means the researcher is unsure exactly how the model makes the classification based on the input [2,5,6]. In the case of an MRI-based diagnosis, the developer is unable to show what regions of the scan the model is using to make its classification. This can cause problems in the diagnosis process, as the medical professional is unable to verify the diagnosis or see the logic behind the AI's classification. As such, there must be a way for any AI designed to diagnose dementia to identify the regions of the MRI that it used to classify its level of dementia.

In this paper, an AI model that addressed both of the previously mentioned issues was developed to classify an MRI scan into one of four categories of dementia. The model also uses explainable AI (XAI) to identify regions of interest in the input scan. The model was built and trained using an open-source database of 6400 MRI scans. Its structure is that of a convolution neural network (CNN), which in the end had a validation accuracy of 98%. Furthermore, the system uses guided backpropagation to highlight the regions of interest in the scan so medical professionals can verify the classification and explain to the patient why they received such a diagnosis. Overall, this combination of CNN and XAI produced a system that could be used in a real-world setting to aid in the diagnosis of dementia so those afflicted can be properly treated.

T. Morris is with the Department of Mechanical, Aerospace, and Biomedical Engineering, University of Tennessee - Knoxville, Knoxville, TN 37916 USA. (e-mail: tmorri35@vols.utk.edu).

Z. Liu is with the Department of Mechanical, Aerospace, and Biomedical Engineering, University of Tennessee - Knoxville, Knoxville, TN 37916 USA. (e-mail: zliu68@vols.utk.edu).

Longjian Liu is with the Department of Epidemiology and Biostatistics, Drexel University, Philadelphia, 19104 PA, USA. (e-mail: ll85@drexel.edu).

X. Zhao is with the Department of Mechanical, Aerospace, and Biomedical Engineering, University of Tennessee - Knoxville, Knoxville, TN 37916 USA. (e-mail: xzhao9@utk.edu).

## II. METHODS

### A. Dataset for Training

The open-source dataset used to build and train the CNN comes from the website *Kaggle* and was uploaded by Sachin Kumar and their collaborator Dr. Sourabh Shastri. The database was last updated in early 2022 and was built to design and develop an accurate framework to classify dementia [7]. The MRI scans were gathered from a variety of other open sources, including the Alzheimer's Disease Neuroimaging Initiative, Data.gov, Alzheimers.net, and the Community Research and Development Information Service. The set contains a total of 6400 MRI scans split among 4 classes: 3200 non-demented, 2240 very mild demented, 896 mild demented, and 64 moderately demented. From this, 80% of the images (2560 non-demented, 1792 very mild demented, 716 mild demented, and 51 moderately demented) were used to train the CNN, and the rest were used for validation.

### B. Convolution Neural Network

The model used in this project is a convolution neural network (CNN), which was chosen due to its ability to process grid-like data (in this case images of MRI scans) and capture features of the images on different spatial scales. Through the use of its convolution filters, the CNN could extract more abstract and complex features than any other classification model. Furthermore, unlike simpler algorithms like support vector machines, the CNN could be designed to output one of four classes instead of binary output. As such, a CNN model was the best fit for this project.

The model was built using the *TensorFlow* and *Keras* packages in Python 3.11. The CNN used in this project included 10 layers: an input layer, 4 convolution layers, a flattening layer, 2 fully connected layers, a dropout layer, and an output layer. The convolution layers used the rectified linear unit (ReLU) activation function to introduce non-linearity, which is a very common activation function for computer-vision-based models. For the convolution layers, the first had 32 filters, the second had 64 filters, the third had 128 filters, and the fourth had 64 filters. Many different structures were tested, including ones with different amounts of convolution layers, dropout, and fully connected layers. The final model is the one that had the highest validation accuracy.

### C. Explainable AI

To aid in the diagnostic process, an explainable AI (XAI) system was developed to identify the regions of the MRI scan that the CNN is using to classify the image. XAIs are used to have models more easily understandable by explaining their output [8]. In the case of this model, an XAI algorithm was used in tandem with the CNN to highlight the pixels of the input image that carry the most weight in its dementia classification. The highlighted pixels are then put back into an image format resembling the original image, which is then used to aid in the diagnostic procedure.

There are many different ways to create an XAI model. This project utilized the *pyplot* library in Python to visualize the filters and outputs from each convolution layer of the CNN. By first visualizing the filters that affect the red, green, and blue pixel values in the RGB image, it can be seen how the CNN determines the weights of each value within a pixel and how those weights change between the different filters in each convolution layer [9]. Furthermore, with this method, the feature maps for each convolution layer can be visualized. This allows users to determine what features of the input image are utilized for classification within each layer [9]. With this method, it is expected that earlier layers detect smaller, more fine-tuned features whereas later layers capture more general features of the input image [9].

### D. Combined System

To fully aid in the diagnosis of dementia, the trained CNN and XAI are combined and used in tandem with a medical professional's expertise. The trained model can be loaded along with the XAI algorithm, then an input MRI can be fed into it. The system will then output 2 things: a dementia classification and an image with the highlighted regions of interest. This system can be seen in **Fig. 1**.

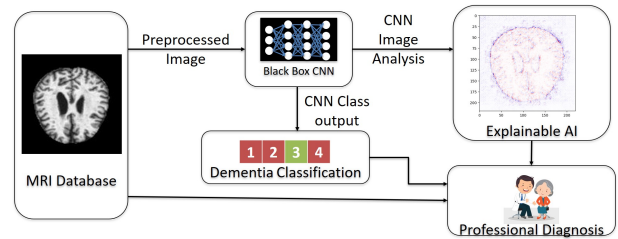

Figure 1. Combination of the CNN and XAI.

## III. RESULTS

After training the CNN for 20 epochs and a batch size of 32, the model had a training accuracy was 99.96% and a training loss was 0.0050. Its validation accuracy and loss were 98.05% and 0.0644, respectively. Furthermore, in terms of individual class accuracies, the model performed well across all classes. As shown by the confusion matrix in **Fig. 2**, the CNN was most accurate in classifying the non-demented class with an accuracy of 99%. The least accurate was moderate dementia with an accuracy of 88%. However, it is notable that all of the moderate dementia MRIs that were misclassified were instead classified as very mild dementia. Regardless, this high level of accuracy asserts this model's feasibility in aiding in the diagnosis of dementia through MRI scans.

In addition to this, K-fold cross-validation was run to further validate the CNN. Using the stratified K-fold toolbox from *sklearn*, the cross-validation was performed with 5 splits, i.e. the database was split into 5 equal parts that represented the distribution of MRI scans among the different classes. After 5 runs, with a different data-part being used for validation and the rest for training, the average validation accuracy was 87% with a standard deviation of 12%.

With the CNN trained and producing highly accurate results, the XAI could be developed. First, the first 6 filters of the first convolution layer were visualized using *pyplot*. The output, as seen in **Fig. 3**, shows the filters in each row, with each column representing the red, green, and blue values respectively. In this visualization, the darker squares indicate a smaller weight in that region of the image and the lighter ones indicate a higher weight. It is worth noting that the filters differ across the RGB values in every layer, indicating that the CNN never weighs the three values equally at any given point.

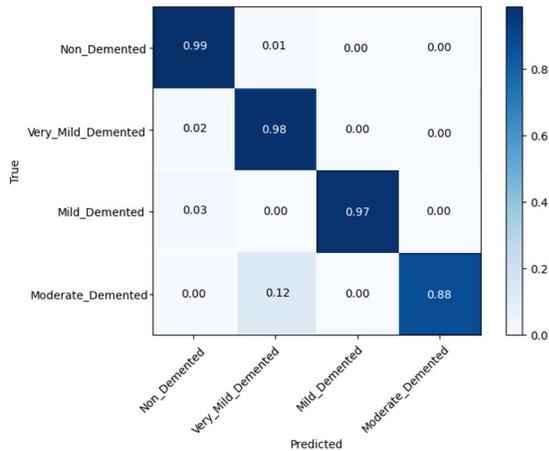

Figure 2. Confusion Matrix for the CNN showing the accuracy for each class.

Another use of the XAI is to visualize the outputs for each of the four convolution layers that make up the model through the generation of feature maps. Again, using *pyplot*, the feature maps were generated for each filter for each layer for example moderate dementia MRI scan. The outputs for the first 6 filters in each of the 4 convolution layers can be seen in **Fig. 4**. Again, the white regions indicate a larger weight and the darker regions indicate a lower weight.

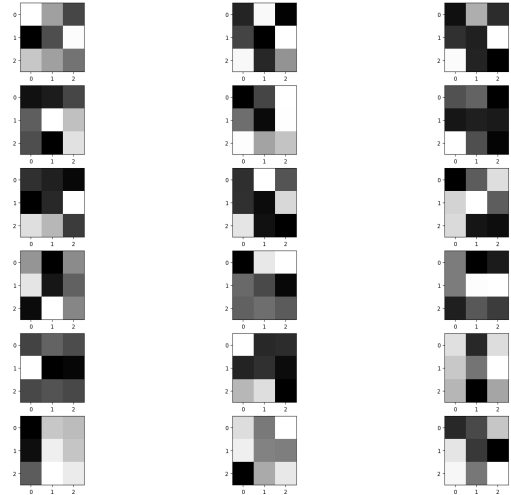

Figure 3. Visualization of the first 6 filters (rows) in the first convolution layer for the red (left), green (middle), and blue (right) values for each pixel.

## IV. Discussion

Looking back on the results of the CNN, the high levels of accuracy and low losses for both the training and validation sets show that the model is a very well fit for the data. The resulting accuracies and losses mean the CNN is more accurate than previous models, which only reached accuracies of up to 96% [2-6]. The small difference between the training and validation sets (both in terms of accuracy and loss) shows that the model is not overfitted to the training data and can therefore generalize well. This is a key component in applying the model in a real-world scenario. By showing that the model not only surpasses other similar models but can also accurately classify new MRIs, we prove that the CNN alone can be adapted in a clinical setting to aid in the diagnosis of dementia through MRI scans.

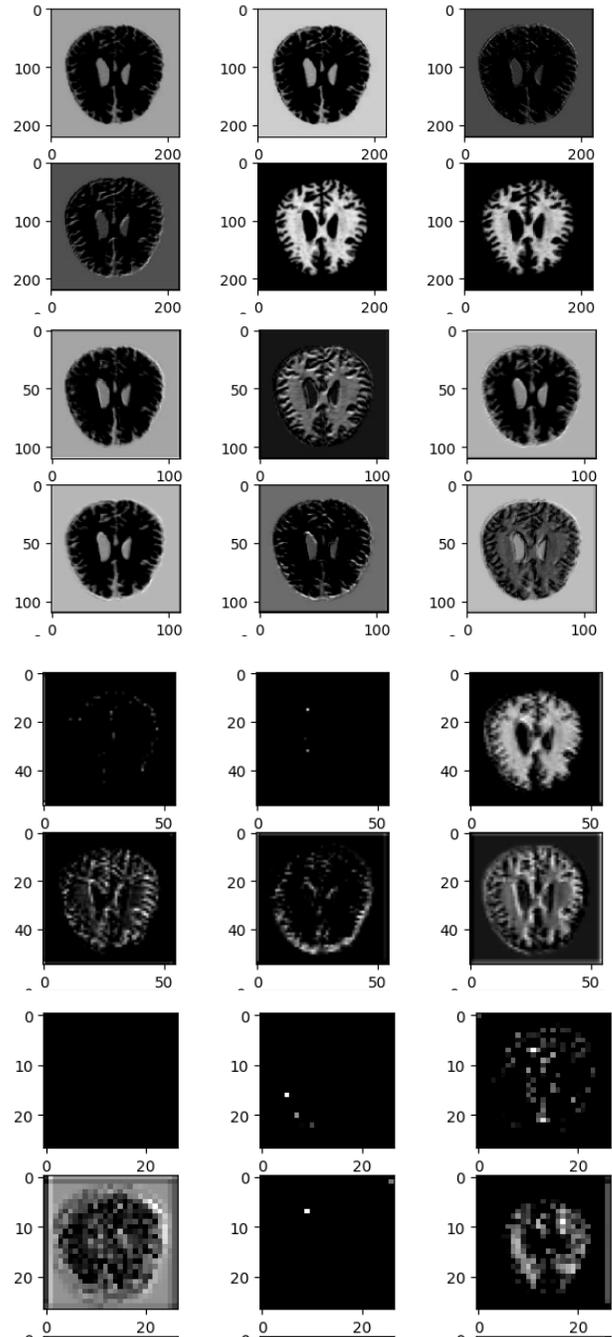

Figure 4. Output of the first 6 filters for each of the 4 convolution layers.

One small issue to note about the CNN is its misclassification of moderate dementia as very mild approximately 12% of the time, as shown in **Fig. 2**. There are a variety of factors that could cause this, the most probable being similarities in the MRIs of very mild and moderate dementia patients. Just like a trained medical professional, the AI is not perfect and can make mistakes if two cases are similar enough. This issue, however, could be fixed with further training and/or additional training methods.

Lastly, considering the high level of accuracy presented by the K-fold cross-validation, the model works well regardless of the validation set and can therefore generalize well. Combining this with the results from the initial training and confusion matrix, it can be concluded that the CNN can be accurately trained and validated using different MRI datasets, as well as applied in a real-world application with a high level of accuracy.

While the accuracy of the CNN is high, there are a few areas of improvements. First, its outputs are absolute, and do not contain a scaled index from non-demented to moderate dementia. Adding such a index would further aid in the diagnosis of a patient. Furthermore, using external data to further validate the model would prove useful in further supporting its accuracy. This, however, could not be done at this time due to external datasets being unavailable. Lastly, the MRI scans contained no demographic data, such as age or sex. These could lead to differences in the MRI scans, which could in turn lead to differences in the model's accuracy. Overall, moving forward, a new dataset that includes more demographic data would enable the CNN to be more accurate and robust in the diagnosis of dementia via MRI scans.

Moving on to the results of the XAI algorithm, the filters in **Fig. 3** show that within different filters, the RGB values can be weighed very differently or similarly. The first filter, for example, puts more weight on the red and green values for each pixel as opposed to the blue values. Furthermore, this figure also highlights the gradients identified within each filter across different regions of the input image, the most prominent being the gradient in the red values in the $6^{th}$ filter. Here, this is a descending gradient from the lower center of the image to the upper right corner, with the left side being ignored for the most part.

The second use of the XAI model pertains more to the diagnostic process. The feature maps generated in **Fig. 4** highlight the regions of the input MRI scan that are most heavily weighted when classifying the stage of dementia. Breaking down the different convolution layers, the feature maps for the first 6 filters in each layer show how each layer captures less fine-grained features and more general features. This can be seen as the first convolution layer having very detailed filter outputs, looking very similar to the original input image. On the other hand, the last convolution layer shows grainy outputs with each filter, indicating that it is identifying larger, more generalized features. These outputs, both the fine-tuned and generalized feature maps, can help medical professionals validate the CNN's classification by cross-examining what it identified as important regions with what they were trained to identify. Overall, this XAI model shows how the CNN works and what image features it uses in its classification as well as helps medical professionals confirm its decision and make a proper diagnosis.

There are issues with this XAI model, however, primarily regarding its inexact highlighting of image features. While the highlighted regions are important and do properly visualize the primary features used in the CNN's classification, an XAI that could produce a sharply outlined region would be easier to interpret; such output could be done by techniques like Layer-wise Relevance Propagation (LRP) or Principal Component Analysis (PCA). Such algorithms, however, would either require the CNN to be restructured and retrained, which might reduce its accuracy, or require more computational power and time, which was not available at the time. Regardless, this XAI algorithm is a great assistant in the understanding of such a complex AI model, as allows both the medical professional and the patient to easily visualize and interpret the results of the MRI scan and understand the reasoning behind the model's classification.

## V. Conclusion

Overall, this combination of a high-level CNN and an XAI algorithm has produced a system that is highly accurate in diagnosing and classifying dementia based on a patient's MRI scan as well as helping the patient and medical professional understand the reasoning behind the classification. The 98% accurate, well-fitted, and unbiased CNN is perfect for analyzing MRI scans and is unlikely to make the same mistakes that a human evaluation might. Furthermore, the XAI helps medical professionals validate the results through the visualization of the CNN's classification process. It also offers an easier-to-understand explanation of the diagnosis/scans to the patient who might be unfamiliar with this kind of medical testing. Overall, this system offers not only a way to accurately diagnose the potential stage of a person's dementia but also explains why the model made the decision that it did.

## Acknowledgment

This work in part is supported by National Institute of Health (NIH) under the grant number R01AG077003.